\documentclass[12pt]{article}
\usepackage[english]{babel}
\usepackage{amsmath,amsthm}
\usepackage{amsfonts}
\usepackage{microtype}
\usepackage{color}
\usepackage{hyperref}
\usepackage{lineno}
\makeatletter
\newif\if@restonecol
\makeatother

\usepackage[linesnumbered,ruled,vlined]{algorithm2e}
\usepackage{algpseudocode}
\usepackage{amsmath}
\usepackage{cite}
\usepackage{listings}
\usepackage{textcomp}
\usepackage{xcolor}
\usepackage[justification=centering]{caption}
\usepackage{graphicx}
\usepackage{array}
\usepackage[braket]{qcircuit}
\usepackage{authblk}

\DeclareCaptionFormat{mylst}{\hrule#1#2#3}
\title{An efficient quantum circuits optimizing scheme compared with QISKit}
\author[1,2]{Xin Zhang}
\author[1,2]{Hong Xiang\thanks{Corresponding author: xianghong@cqu.edu.cn}}
\author[2,3]{Tao Xiang}
\author[1,2]{Li Fu}
\author[1,2]{Jun Sang}
\affil[1]{School of Big Data and Software Engineering, Chongqing University}
\affil[2]{Key Laboratory of Dependable Service Computing in Cyber Physical Society (Chongqing University), Ministry of Education}
\affil[3]{School of Computer Science, Chongqing University}
\date{July 3, 2018}%
\begin{document}
\maketitle
\begin{abstract}
Recently, the development of quantum chips has made great progress-- the number of qubits is increasing and the fidelity is getting higher. However, qubits of these chips are not always fully connected, which sets additional barriers for implementing quantum algorithms and programming quantum programs. In this paper, we introduce a general circuit optimizing scheme, which can efficiently adjust and optimize quantum circuits according to arbitrary given qubits' layout by adding additional quantum gates, exchanging qubits and merging single-qubit gates. Compared with the optimizing algorithm of IBM's QISKit, the quantum gates consumed by our scheme is 74.7\%, and the execution time is only 12.9\% on average.
\end{abstract}
\section{Introduction}
Quantum computing has attracted increasing attention because of its tremendous computing power \cite{simon1997power,shor1999polynomial,grover1996fast} in recent years. There are more and more companies and scientific research institutions  who devote themselves to developing quantum chips with more qubits and higher fidelity. While most theoretical studies assume that interactions between arbitrary pairs of qubits are available, almost all these realistic chips have certain constraints on qubit connectivity\cite{cheung2007translation,linke2017experimental}.
For example, IBM's 5-qubit superconducting chips \textit{Tenerife} and \textit{Yorktown}\cite{ibmbi} adopt neighboring connectivity ( illustrated in Fig.1 (a) and 1 (b), respectively). \cite{zhong2016emulating} uses a 4-qubit superconducting chip, in which four qubits are not directly connected, but are connected by a central resonator. That is, the layout of this chip is central, as shown in Fig.1 (c). In addition, CAS-Alibaba Quantum Laboratory's 11-qubit superconducting chip\cite{alibaba} and Tsinghua University's 4-qubit NMR chip\cite{xin2017nmrcloudq} both reduce the fully connectivity to the linear nearest-neighbor connectivity, as shown in Fig.1 (d). Distinctly, this non-fully connected connection sets additional barriers for implementing quantum algorithms and programming quantum programs.
 \begin{figure}[!htbp]
    \begin{minipage}[t]{0.24\linewidth}
     \centering
 	  \includegraphics[width=0.8\textwidth]{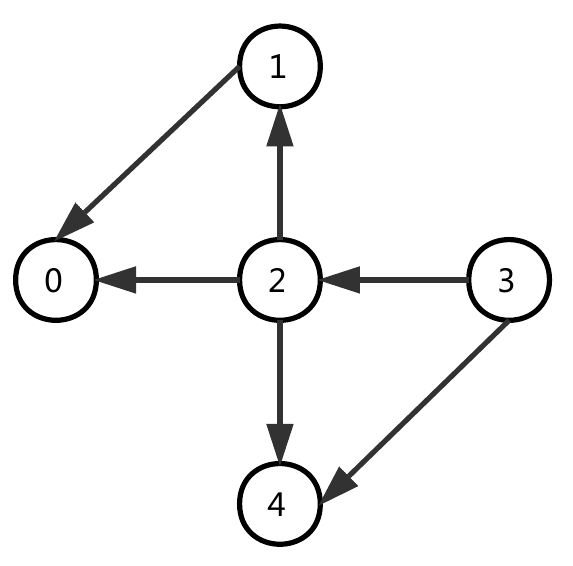}
 	  \caption*{(a) Tenerife}
    \end{minipage}
 	\begin{minipage}[t]{0.24\linewidth}
     \centering
 	  \includegraphics[width=0.8\textwidth]{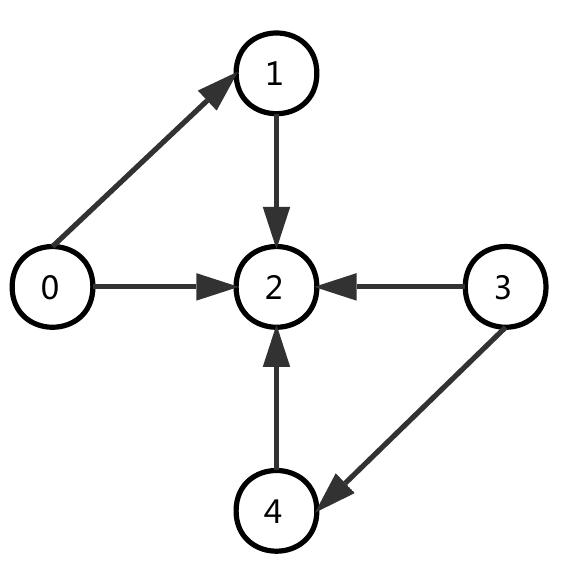}
 	  \caption*{(b) Yorktown}
    \end{minipage}
 	\begin{minipage}[t]{0.24\linewidth}
     \centering
 	  \includegraphics[width=0.8\textwidth]{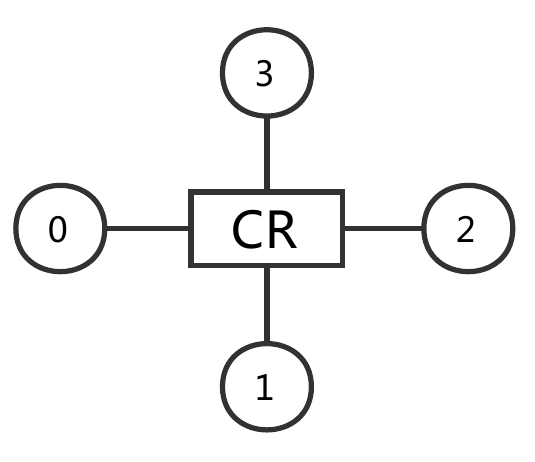}
 	  \caption*{(c) Central layout}
    \end{minipage}
 	\begin{minipage}[t]{0.24\linewidth}
     \centering
 	  \includegraphics[width=0.8\textwidth]{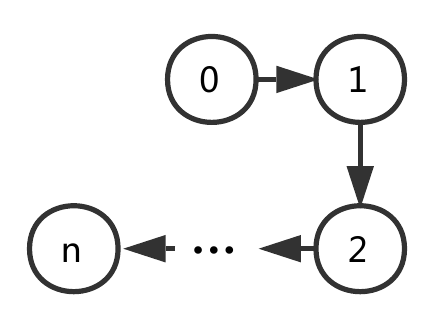}
 	  \caption*{(d) Linear layout}
    \end{minipage}
    \caption{The four different physical layouts}
 \end{figure}
\\
\indent On the other hand, decoherence\cite{zeh1970interpretation} is a huge challenge for quantum computing and the quantum programs should be executed within coherence time\cite{divincenzo2000physical}. For getting more reliable results, we need to reduce the quantum circuit depth\cite{yao1993quantum} as far as possible. However, for non-fully connected physical layouts, if we want to execute arbitrary quantum programs, we must add additional quantum gates to adjust the original quantum program, which will inevitably lead to an increase in the depth. Therefore, it is of great practical significance to design an optimization algorithm which can minimize the overhead as mush as possible.\\
\indent As early as 2007, D. Cheung et al. made a discussion about the non-fully connected physical layout\cite{cheung2007translation}. By adding SWAP gates, they turned illegal CNOT operations into legitimate operations and proved that the star-shaped or the linear nearest-neighbor connectivity could be able to utilize additional $O(n)$ quantum gates to complete the adjustment, where $n$ stands for the number of qubits. In 2017, IBM developed a quantum information science kit, namely QISKit \cite{qiskit}, which contains an algorithm that can adjust and optimize quantum programs according to any layout. Recently, in order to find more efficient solutions, IBM organized the QISKit Developer Challenge \cite{qiskitc}. As for the optimization of quantum circuits, in order to simulate more qubits on classical computers, E. Pednault et al. proposed a method, namely \textit{slice}\cite{pednault2017breaking}, to split the original quantum circuit into multiple subcircuits. In this way, they simulate a random quantum circuit with depth 27 in a 2D lattice of $7\times7$ qubits and a circuit with depth 23 in a 2D lattice of $8\times 7$ qubits on the IBM Blue Gene/Q supercomputer, which improved the number of entangled qubits that classical computers can simulate. However, the \textit{slice} approach is focused on the simulation of more entangled qubits, so it do not take into account the physical layout, and is only applicable to programs with short circuit depth. \\
\indent In this paper, we propose a general enough quantum circuit optimizing scheme which can efficiently adjust and optimize any quantum circuit according to any layout. The remainder of this paper is organized as follows: Section 2 briefly introduce the necessary conceptions. In Section 3, the design concept of our optimizing scheme is presented in detail. We next compare the cost and efficiency of our scheme with QISKit's optimizing method in Section 4. The conclusion and future research can be found in Section 5. 
\section{Preliminaries}

\subsection{QISKit}
QISKit is a quantum information science kit developed by IBM, which takes the quantum programs written by Open-QASM\cite{cross2017open} as the input. It adjusts and optimizes the input programs according to the given layout, and then executed the programs by its built-in QASM-simulator or cloud-based quantum chips.\\
\indent Open-QASM is a variant of QASM\cite{svore2006layered}, which is designed to control a physical system with a parameterized gate set. Specifically, Open-QASM takes $\{ u1,u2,u3,CNOT \}$ as the basic quantum gates set, where
\begin{align*}
&u1(\lambda) = \left( \begin{array}{cc} 1&0\\0&e^{\lambda i} \end{array} \right),\\
&u2(\phi,\lambda) =\frac{1}{\sqrt{2}} \left( \begin{array}{cc} 1 & -e^{\lambda i}\\ e^{\phi i} & e^{(\lambda i+\phi i)} \end{array} \right),\tag{1}\\
&u3(\theta,\phi,\lambda) = \left( \begin{array}{cc} \cos{\frac{\theta}{2}}&-e^{\lambda i}\sin{\frac{\theta}{2}}\\ e^{\phi i}\sin{\frac{\theta}{2}}&e^{(\lambda i+\phi i)}\cos{\frac{\theta}{2}} \end{array} \right).
\end{align*}
Obviously, $\{ u1,u2,u3,CNOT \}$ actually has an infinite number of single-qubit gates and it is universal\cite{Barenco1995Elementary}. For comparison with QISKit, our optimizing scheme also takes it as the basic set of quantum gates.
\subsection{Common solutions}
\indent Before introducing the common solutions, we need to point out the main obstacles for hindering the execution of quantum programs:
\begin{itemize}
\item Obstacle-1: the direction of CNOT gate is illegal, as shown in the red line in Fig.2 (a);
\item Obstacle-2: the connectivity between two specific qubits is illegal, as shown in the blue line.
\end{itemize}
 \begin{figure}[!htbp]
    \begin{minipage}[t]{0.48\linewidth}
     \centering
 	  \includegraphics[width=0.4\textwidth]{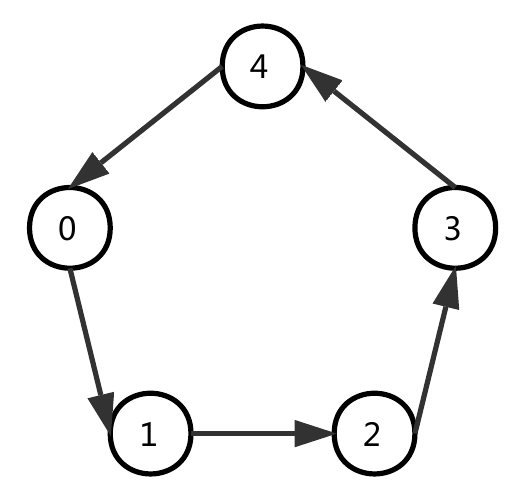}
 	  \caption*{(a) Given Layout}
    \end{minipage}
 	\begin{minipage}[t]{0.48\linewidth}
     \centering
 	  \includegraphics[width=0.4\textwidth]{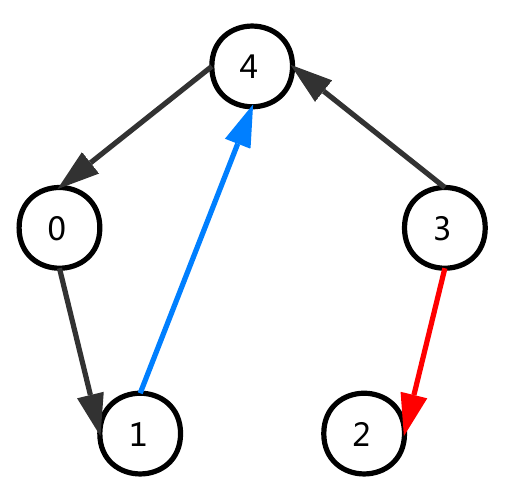}
 	  \caption*{(b) Actual Layout}
    \end{minipage}
    \caption{An example of Obstacle-1 and Obstacle-2.}
 \end{figure}
\indent For Obstacle-1, a common solution is to flip the direction by 4 additional H gates:
$$
\text{H}_2 \times \text{CNOT}_{(q_1,~q_2)} \times \text{H}_2 = \text{CNOT}_{(q_2,~q_1)}.\eqno{(2)}
$$
As for Obstacle-2, the basic idea is exchanging the states of qubits by SWAP gates. For example, although \textit{cnot($q_1$, $q_4$)} is illegal in Fig.2 (a), we can use another way to accomplish the same task, such as the circuit shown in Fig.3.
\\
\indent However, the additional overhead of this solution is costly, especially for sparse physical layouts. Specifically,
$$
cost = 2m \times cost_{\text{SWAP}},\eqno{(3)}
$$
where $m$ stands for the number of intermediate nodes on the shortest path between the control-qubit and the target-qubit, $cost_{\text{SWAP}}$ stands for 3 CNOT gates and 4 H gates.
 \begin{figure}[!htbp]
    \begin{minipage}[t]{0.5\linewidth}
     \centering
 	  \includegraphics[width=0.6\textwidth]{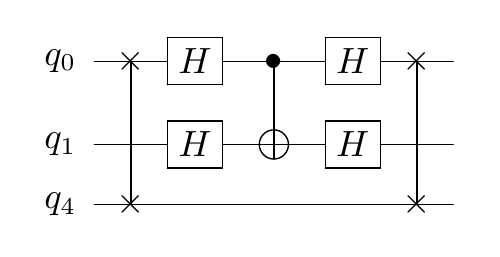}
 	  \caption*{(a) An implementation of \textit{cnot($q_1$,~$q_4$)}}
    \end{minipage}\hfill
 	\begin{minipage}[t]{0.5\linewidth}
     \centering
 	  \includegraphics[width=0.6\textwidth]{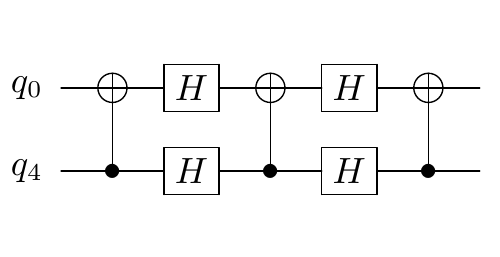}
 	  \caption*{(b) \textit{SWAP($q_0$,~$q_4$)}}
    \end{minipage}
    \caption{An equivalent circuit of {cnot($q_1$,~$q_4$)}, where \textit{SWAP($q_0$,~$q_4$)} is implemented by (b).}
 \end{figure}

\section{Our Optimizing Scheme}
As mentioned before, the non-fully connected layout is widely adopted. There are only two ways to execute arbitrary quantum programs:
\begin{itemize}
\item Hardware solution: Completely changing the layouts of chips and constructing fully connected chips;
\item Software solution: Designing a circuit optimization algorithm, which is able to adjust the original quantum program to meet requirements of the chip.
\end{itemize}
\indent Our optimizing scheme is an efficient general solution from software level. Specifically, we design the following three steps to adjust and optimize quantum programs based on the common solutions described in Section 2.2.
\subsection{The global adjustment of qubits}
The global adjustment of qubits means that before the execution of quantum programs, we first compare the connected relation of quantum programs with the given layout, and directly exchange the qubits.
The greatest advantage of this step is that no additional quantum gates need to be consumed.
Therefore, the number of additional quantum gates consumed will be minimum if all illegal CNOT gates can be handled in this step.
For simplicity, we assume that any edge in the given layout is bidirectional in this step and \textit{Local adjustment}, that is, the Obstacle-1 is ignored in the two steps.\\
\indent Specifically, this step can be described as Algorithm 1.
In Algorithm 1, we extract all CNOT gates from the quantum program separately and traverse them from front to back.
Once encountering an illegal CNOT gate, we try to find an available qubits' mapping to adjust the whole Open-QASM code without converting the traversed CNOT gates illegal.
At each adjustment, we have $(d_{cq}\times d_{tq}-t)$ available mappings to choose, where $t$ stands for the number of mappings which make some traversed CNOT gates illegal, $d_{cq}$ and $d_{tq}$ stand for the number of adjacent qubits of control-qubit and target-qubit in the given layout, respectively.
The traversal terminates when there is no illegal CNOT gate or $(d_{cq}\times d_{tq}-t)=0$.\\
\indent Suppose that there are $M$ possible mappings, where $M$ is related to the given layout and the connectivity of quantum programs.
At this point, we need to estimate the cost of solving Obstacle-2 in the program adjusted according to these $(M+1)$ mappings ($M$ mappings and one empty mapping) respectively. Then take the smallest one as the global adjustment mapping. The reason for estimation, rather than accurate calculation, and the estimation process are explained in the next part.
Finally, we adjust the qubits of the original Open-QASM code according to the global mapping. As for the classical register, which stores the results of the measurement, does not need to be modified.
For example, $cnot(q_1,~q_4)$ is illegal in Fig.2 and it can be adjusted by the global mapping $\{1:3,~3:1\}$, as shown in Fig.4.

 \begin{algorithm}
  \caption{Global Adjustment}
  \KwIn{The set of CNOT in QP, $C$; the set of legal CNOT, $A$; the record of all possible costs, $costs$; the record of all possible mappings, $maps$; the current mapping, $amap$;}
  \KwOut{The mapping of qubits' ID, $map$}
  \SetKwFunction{funone}{GlobalAdjust}
  \SetKwBlock{funones}{\funone{$costs$,~$maps$,~$amap$}}{end}
  \SetKwFunction{funtwo}{Adjust}
  \SetKwBlock{funtwos}{\funtwo{$C$,~$A$,~$amap$,~$costs$,~$maps$}}{end}
  \funones
  {
        $costs\leftarrow $[ ],$maps\leftarrow$[ ] and $amap\leftarrow$[ ]\;
        Adjust($C$,~$A$,~$amap$,~$costs$,~$maps$)\;
        $i\leftarrow$getIndexofMinValue($costs$)\;
        return $maps[i]$\;
  }
  \quad\\
  \funtwos
  {
	  $alternativeMap$ $\leftarrow$ [ ]\;
	  \For {\rm CNOT $c$ in $C$}
      {
        \If {\rm $c$ not in $A$}
        {
            $cq\leftarrow c[0]$ and $tq\leftarrow c[1]$\;
            $cqAdj$ $\leftarrow$ getAdjacentQubit($cq$) and $tqAdj$ $\leftarrow$ getAdjacentQubit($tq$)\;
            $tMaps\leftarrow\{cq:tqAdj,~tq:cqAdj\}$\;
            \For {\rm map $m$ in $tMaps$}
            {
                $tempC\leftarrow C$\;
                change qubits' ID in $tempC$ according to $m$\;
                \If {\rm no illegal CNOT in $tempC$}
                {
                    add $m$ to $alternativeMap$\;
                }
            }
            \textbf{break}\;
        }
      }
      \If {\rm $alternativeMap$ == [ ]}
      {
        $cost$ $\leftarrow$ estimateCost()\;
        add $cost$ to $costs$ and add $amap$ to $maps$\;
      }
	    \For {\rm map $am$ in $alternativeMap$}
        {
            $tempC \leftarrow C$ and add $am$ to $amap$\;
            change qubits' ID in $tempC$ according to $am$\;
            \If {\rm no illegal CNOT in $tempC$}
	        {
            add $amap$ to $maps$ and add $0$ to $costs$\;
            }
            \Else
            {
            Adjust($C$,~$A$,~$amap$,~$costs$,~$maps$)\;
            }
        }
}
\end{algorithm}

 \begin{figure}[!htbp]
    \begin{minipage}[t]{0.5\linewidth}
     \centering
 	  \includegraphics[width=0.9\textwidth]{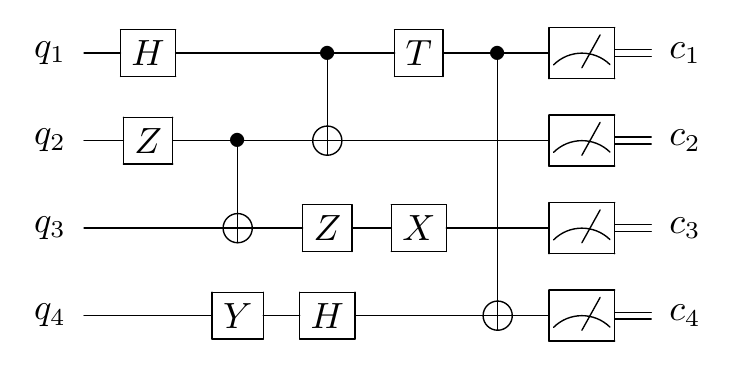}
 	  \caption*{(a) Before adjusting}
    \end{minipage}\hfill
 	\begin{minipage}[t]{0.5\linewidth}
     \centering
 	  \includegraphics[width=0.9\textwidth]{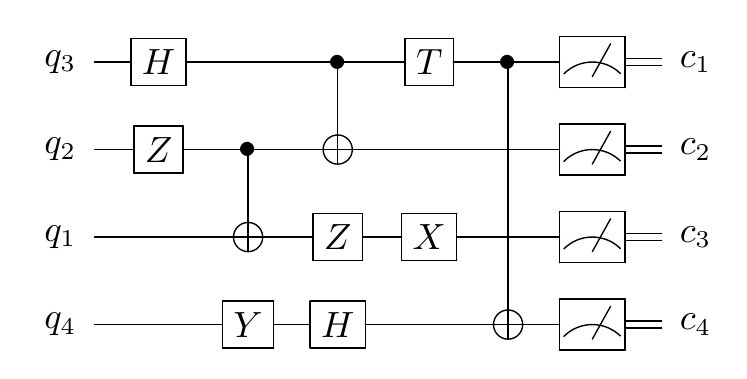}
 	  \caption*{(b) After adjusting}
    \end{minipage}
    \caption{Adjust the circuit according to $\{1:3,~3:1\}$ and (b) can be executed on Fig.2 (a)}
 \end{figure}

\subsection{The local adjustment of qubits}
In this step, the exchange of qubits' state mainly depends on adding SWAP gates.
Compared with the basic solution described in Section 2.2, our scheme has the following differences:
\begin{itemize}
\item There is no need to use SWAP gates again to restore the state. Instead, we use the qubits involved in the exchange and intermediate qubits to generate a local mapping, then modify the subsequent gates and classical registers according to the mapping;
\item Due to the existence of the first difference, the effect of exchanging control-qubit with intermediate qubits by SWAP gates and exchanging target-qubit with these qubits is completely different for the subsequent code. Therefore, we need to calculate the gate costs in the two cases respectively and take the smaller one as the object of exchange.\\
\end{itemize}
\indent ~~~~However, it is difficult to accurately calculate the costs of these two cases in the second difference.
During the calculation, we will encounter several illegal CNOT gates, and for each illegal CNOT, we have two solutions.
Actually, the solution space is a binary tree whose height is $n$ and the number of leaf nodes is approximately $O(2^n)$, where $n$ stands for the number of illegal CNOT gates.
Obviously, classical computers have no ability to complete such large-scale calculations in a relatively short time and we can only estimate the cost. Essentially, the estimation process is based on greedy ideas and easily trapped into the local optimization.
With the increase in the scale of quantum programs, the manifestation of this greedy choice is more obvious, which can be seen in Section 4.\\
\indent In our scheme, the cost of adjusting the Open-QASM code $qasm$ is estimated by
$$cost_{qasm} = \sum_{i=1}^{n}[(\frac{n-i}{n})^2\cdot m_i\cdot cost_{\text{SWAP}}],\eqno{(4)}
$$
where $m_i$ stands for the number of intermediate qubits between the control-qubit and the target-qubit of the $i$th illegal CNOT, and $cost_{\text{SWAP}}$ stands for 3 CNOT gates and 4 H gates.
Among the various estimation formulas we tried, the result obtained by Equation (4) is optimal.
The reason for adding the correction factor $(\frac{n-i}{n})^2$ in Equation (4) is that the later the CNOT gate is executed, the easier it is influenced by the previous adjustments.
That is, estimation is not reliable for the later CNOT gates.
Multiplying the factor, which will continue to decrease as the estimation progress, with the estimation results can have a certain correction effect.\\
\indent For improving the accuracy of estimation, we accurately calculate the top $4$ layers of the binary tree, and estimate the cost of the subsequent gates of the $2^4$ cases respectively, where $4$ is the optimal value determined after repeated trials.
Then add the estimated result and the calculated result together and choose the smallest one among the 16 cases as our choice.\\
\indent Specifically, we traverse the Open-QASM code. Whenever encountering an illegal CNOT, we call Algorithm 2 to adjust it and then update the subsequent code and the classical register until the traversal terminates. It can be seen from Algorithm 2 that the mapping generated by \textit{Adjust} function only affects the subsequent code of $illC$ and that is why we call this step \textit{Local adjustment}.\\
\indent At this point, there is no Obstacle-2 in quantum programs. Then we traverse the new Open-QASM code again to handle Obstacle-1 by Equation (2).
\begin{algorithm}
  \caption{Local Adjustment}
  \KwIn{The Open-QASM code of the quantum program, $qasm$; the first illegal CNOT, $illC$; the rest CNOTs after $illC$ in $qasm$, $Cs$; the record of all possible costs, $costs$; the cost in the current case, $cost$; the record of all possible mappings, $maps$; the current mapping, $amap$; the depth of recursion, $d$}
  \KwOut{The adjusted Open-QASM code, $qasm$}
  \SetKwFunction{funone}{LocalAdjust}
  \SetKwBlock{funones}{\funone{$qasm$,~$illC$,~$Cs$}}{end}
  \SetKwFunction{funtwo}{Adjust}
  \SetKwBlock{funtwos}{\funtwo{$illC$,~$Cs$,~$cost$,~$costs$,~$map$,~$maps$,~$d$}}{end}
    \funones
    {
        $cost\leftarrow 0$, $costs\leftarrow $[ ], $amap\leftarrow$[ ], $d\leftarrow 1$ and $maps\leftarrow $[ ]\;
        Adjust($illC$,~$Cs$,~$cost$,~$costs$,~$amap$,~$maps$,~$d$)\;
        $i\leftarrow$getIndexofMinValue($costs$)\;
        add SWAP gates to $qasm$ according to $maps[i]$\;
        change qubits'ID in $qasm$ according to $maps[i]$\;
        return $qasm$\;
    }
    \quad\\
    \funtwos
    {
        $interQs\leftarrow$ getIntermediateNode($illC[0]$,~$illC[1]$)\;
        $cost\leftarrow cost + 34\times $$interQs$.length\;
        \For {\rm qubit $q$ in $illC$}
        {
            $tc\leftarrow cost$\;
            \If {\rm $q$ is control-qubit}
            {
                $tc\leftarrow tc + 4$\;
            }
            $tMap\leftarrow$ constructMapBetweenQ($interQs$,$q$)\;
            change qubits' ID in $cnots$ according to $tMap$\;
            $nIllC\leftarrow $ getFirstIllegalCnot($cnots$)\;
            $restC\leftarrow$ getAllCnotAfterNewIllC($cnots$)\;
            \If {\rm $map$ != [ ]}
            {
                $tMap\leftarrow map$\;
            }
            \If {\rm $nIllC==$ None}
            {
                add $tc$ to $costs$ and $tMap$ to $maps$\;
            }
            \ElseIf {\rm $d$ == $4$}
            {
                $tc\leftarrow tc$ + estimateCost()\;
                add $tc$ to $costs$ and add $tMap$ to $maps$\;
            }
            \Else
            {
                Adjust($nIllC$,~$restC$,~$tc$,~$costs$,~$tMap$,~$maps$,~$d+1$)\;
            }
        }
    }
\end{algorithm}
\subsection{The mergence of single-qubit gates}
In this step, we will reduce the circuit depth by merging single-qubit gates.
At first, we need to determine which kind of single-qubit gates can be merged.
\\
\indent The random quantum circuit shown in Fig.5 (a) contains three CNOT gates and these gates divide the execution processes of $q_0$, $q_1$, $q_2$ into three parts respectively. Obviously, single-qubit gates in these parts can be merged and we can reduce Fig.5 (a) to Fig.5 (b). Based on this example, we can draw a conclusion that for any qubit $q$, the $n$ multi-qubit gates with $q$ involved can divide the execution process of $q$ into $n+1$ subintervals and the single-qubit gates in each subintervals can be merged into one gate.
 \begin{figure}[!htbp]
    \begin{minipage}[t]{0.5\linewidth}
     \centering
 	  \includegraphics[width=0.9\textwidth]{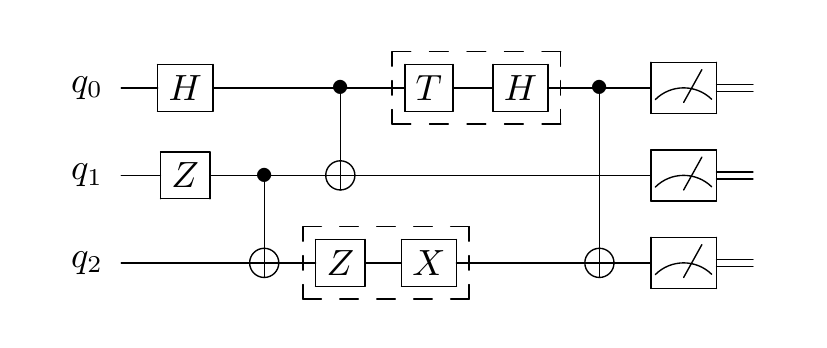}
 	  \caption*{(a) Before merging}
    \end{minipage}\hfill
 	\begin{minipage}[t]{0.5\linewidth}
     \centering
 	  \includegraphics[width=0.9\textwidth]{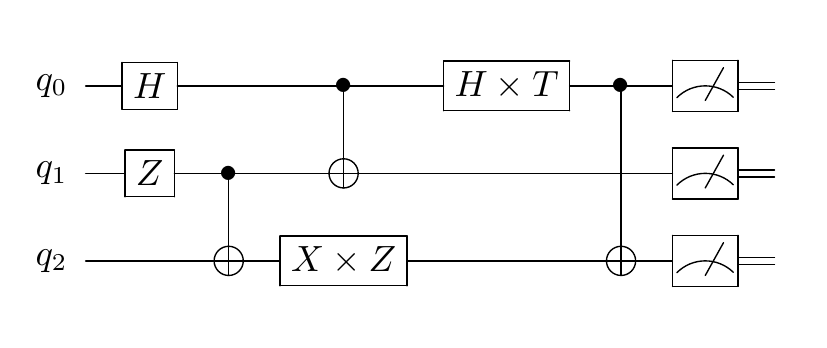}
 	  \caption*{(b) After merging}
    \end{minipage}
    \caption{The change of a quantum random circuit before and after merging single-qubit gates.}
 \end{figure}
\\
\indent As mentioned before, all single-qubit gates in Open-QASM belong to $\{u1,u2,u3\}$. Therefore, merging single-qubit gates actually contains 9 different cases: $u1\times u1$, $u1\times u2$, $u1\times u3$, $u2\times u1$, $u3\times u1$, $u2\times u2$, $u3\times u2$, $u2\times u3$ and $u3\times u3$. In order to handle these cases, we need to do \textbf{Z-Y decompositions}\cite{nielsen2002quantum} for $u1$, $u2$ and $u3$.
By Equations (1), we obtain:
\begin{align*}
&u1(\lambda) = R_z(\lambda),\\
&u2(\phi,\lambda) = u3(\frac{\pi}{2},\phi,\lambda) = R_z(\phi)\times R_y(\frac{\pi}{2}) \times R_z(\lambda),\tag{5}\\
&u3(\theta,\phi,\lambda) = R_z(\phi)\times R_y(\theta) \times R_z(\lambda);
\end{align*}
For the first five cases, we can directly merge them by $R_z(\lambda)\times R_z(\phi) = R_z(\lambda+\phi)$\cite{Barenco1995Elementary}. As for the last four cases, we have:
\begin{align*}
&R_z(\phi_1)\cdot R_y(\theta_1)\cdot R_z(\lambda_1) \cdot R_z(\phi_2)\cdot R_y(\theta_2)\cdot R_z(\lambda_2)\\
&=R_z(\phi_1)\cdot [R_y(\theta_1)\cdot R_z(\lambda_1 + \phi_2)\cdot R_y(\theta_2)]\cdot R_z(\lambda_2)\\
&=R_z(\phi_1)\cdot[R_z(\alpha)\cdot R_y(\beta)\cdot R_z(\gamma)]\cdot R_z(\lambda_2)\tag{6}\\
&=R_z(\phi_1+\alpha)\cdot R_y(\beta)\cdot R_z(\gamma+\lambda_2)\\
&=u3(\beta,\phi_1+\alpha,\gamma+\lambda_2).
\end{align*}
The key of this kind of merging lies in how to transform the \textbf{Y-Z decomposition} of a quantum gate to the \textbf{Z-Y decomposition}. And we use QISKit's merge method proposed in \cite{u3u3} to solve this problem.
So far, we complete the adjustment and optimization of the original quantum program according to any given layout.
\section{Numerical Results}
In this section, we take QISKit's optimizing method as the benchmark to evaluate the performance of our optimizing scheme in different scales of quantum programs and different layouts of quantum chips.
In addition, we use the method proposed in the QISKit Developer Challenge to count the cost of gates:
$$cost=n_2\times 10 + n_1\times 1,\eqno{(7)}
$$
where $n_2$ and $n_1$ stand for the number of CNOT gates and single-qubit gates in optimized quantum circuit, respectively.
\subsection{Platform}
\noindent\textbf{Hardware Platform}\\
\indent All the experiments in this paper are executed on a PC with an Intel Core i7 processor and 8GB of RAM. Furthermore, we have no special hardware acceleration, such as a GPU.\\
\noindent\textbf{Software Platform}\\
\indent In order to verify the correctness of our scheme, we use the QASM-simulator to execute the optimized circuits. In addition, we also use a special method to generate random quantum circuits, which first generates random circuits whose quantum gates belong to $SU(4)$ \cite{iachello2006lie}, and then decomposes these gates into gates belonged to $\{SU(2),~CNOT\}$ \cite{vatan2004optimal}. The advantage of this method is that we can fully test different connections between qubits and the fairness of comparison between our optimizing scheme and QISKit (version=0.4.11) can be guaranteed.
The detailed execution flow of our experiments is shown in Fig.6.
 \begin{figure}[!htbp]
 	\centering
 	\includegraphics[width=0.4\textwidth]{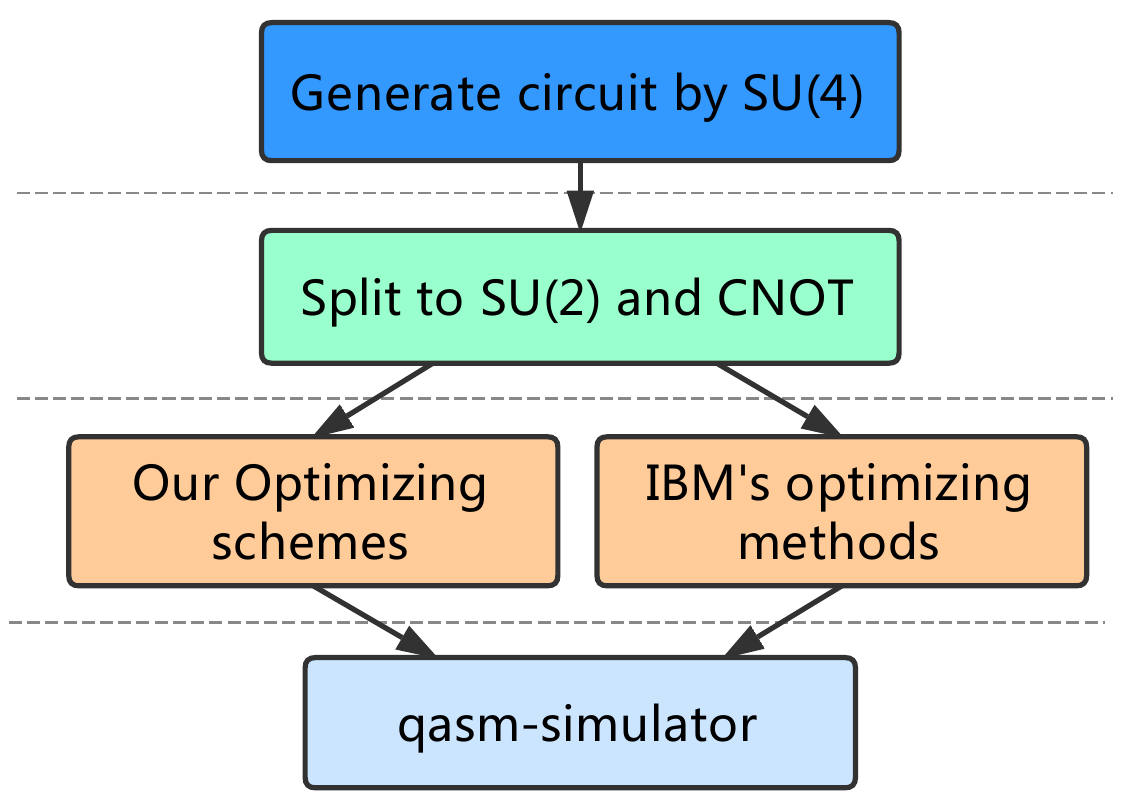}
 	\caption{Execution Flow Chart}
 	\label{figure.3}
 \end{figure}
It should be noticed that for accurate description, the circuit depth mentioned in the following is still $SU(4)$ circuit's depth, and the actual depth is about
7 times of it.
\subsection{Results}
As we all know, the number of qubits and the circuit depth are important indicators for the scale of quantum programs.
Therefore, the experiments are designed as follow: for the $14$ cases of qubits number from $3$ to $16$, we generate $10$ different random quantum circuits  respectively for $16$ cases with circuit depth from $1$ to $16$ respectively. That means, in total, $14\times 16\times10=2240$ circuits are generated.
Then we chose four common connected graphs (linear, central, neighboring and circular) and use our optimizing scheme and QISKit's algorithm to adjust and optimize these $2240$ random circuits according to these layouts, respectively. That is, each algorithm handles $8960$ ($2240\times 4$) quantum circuits.
Finally, the optimized quantum programs are executed by QASM-simulator. If the result of our scheme is consistent with QISKit's result, we count the cost and the execution time of each circuit.\\
\indent All quantum circuits, layouts and the source code of our scheme can be found in Github\footnote{\url{https://github.com/zhangxin20121923/QISKit_Deve_Challenge}}.\\
\noindent\textbf{Comparison with QISKit's optimizing method}\\
\indent Table 1 shows the quantum gates consumption of the $2240$ original random quantum circuits, and the average cost of gates and compiler time required to adjust and optimize these $2240$ circuits by our scheme and QISKit.
\begin{table}[!htbp]
\caption{The overall statistical}
\begin{center}
\begin{tabular}{|p{3.2cm}<{\centering}|p{2cm}<{\centering}|p{2cm}<{\centering}|}
\hline
&Time (s)&Gate Cost\\ \hline
Original Circuit&0 & 3084391\\ \hline
Our Scheme&16472.48 &6703061\\ \hline
QISKit&127751.99 &8974717 \\ \hline
\end{tabular}
\end{center}
\end{table}
Obviously, the quantum gates consumed by our scheme is 74.7\% of QISKit, and the execution time is only 12.9\%.\\
\indent Specifically, the performance of our scheme varies for different scales of quantum circuits.
 \begin{figure}[!htbp]
    \begin{minipage}[t]{0.45\linewidth}
     \centering
 	  \includegraphics[width=\textwidth]{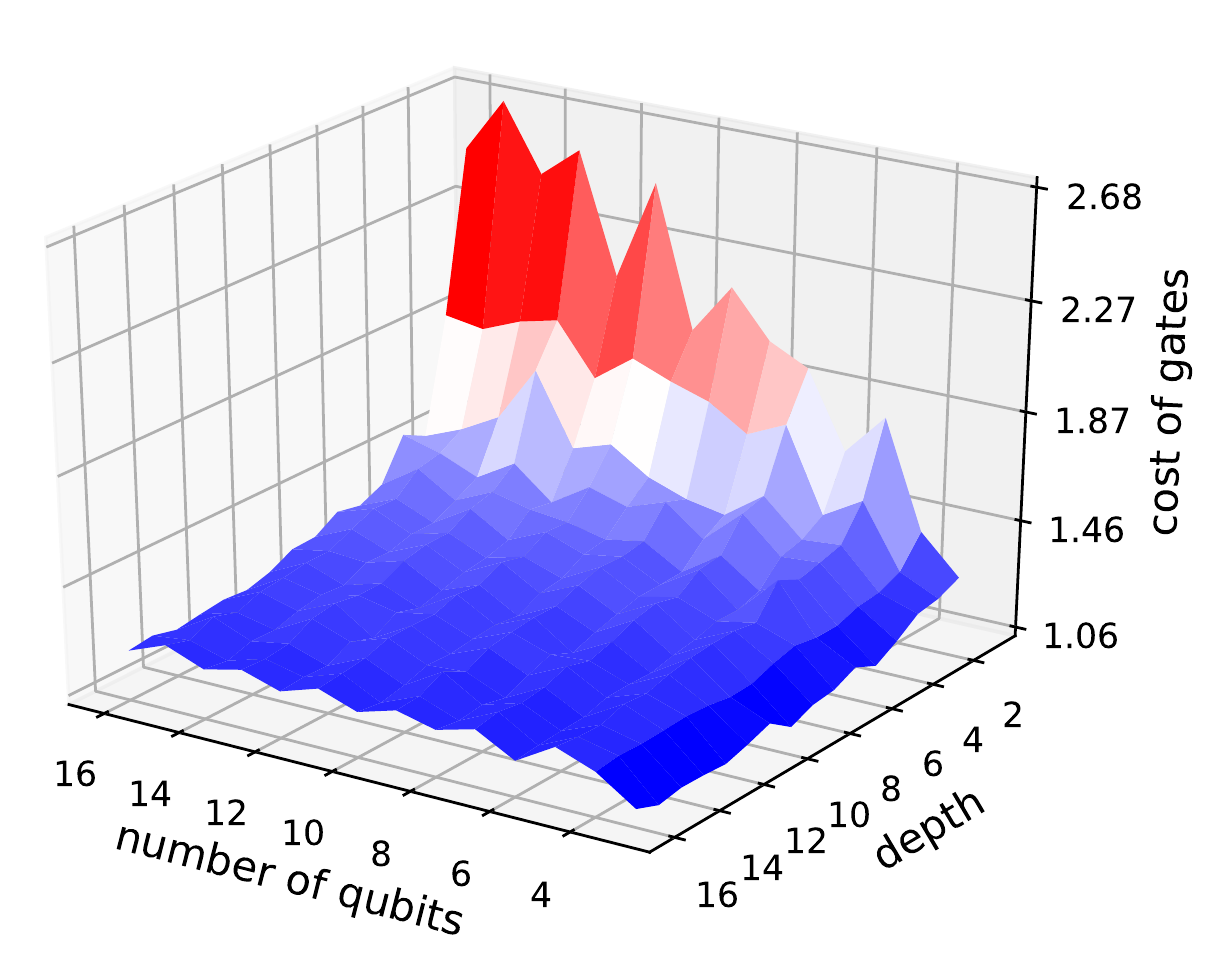}
 	  \caption*{(a) Gate Cost}
    \end{minipage}\hfill
 	\begin{minipage}[t]{0.45\linewidth}
     \centering
 	  \includegraphics[width=\textwidth]{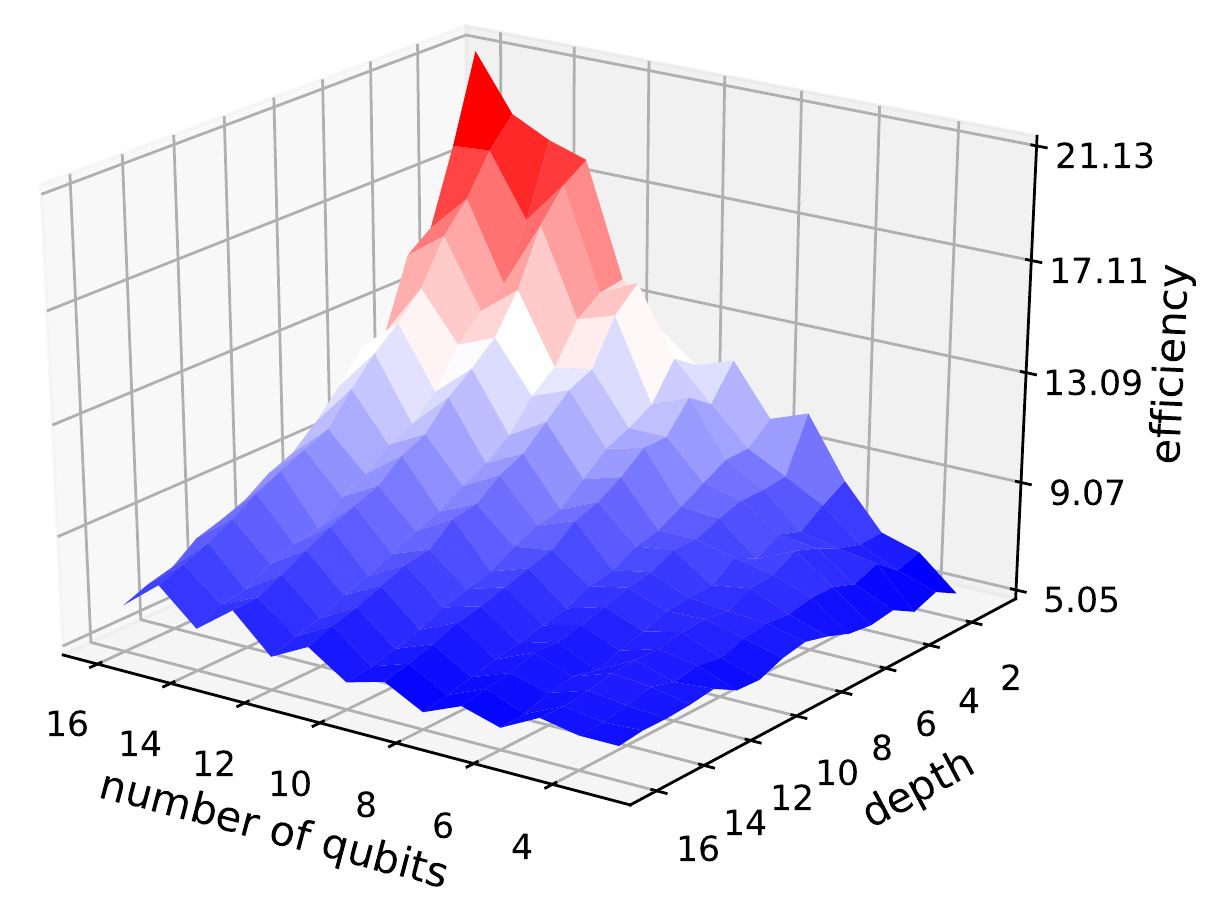}
 	  \caption*{(b) Efficiency}
    \end{minipage}
    \caption{Experimental Results}
 \end{figure}
Fig.7 (a) and Fig.7 (b) illustrate the ratio of QISKit and our scheme about the cost of quantum gates and efficiency
with various qubits $q$ and circuit depths $d$, respectively. The two formulas are shown as follows:
$$\text{cost}_{(n,d)} = \frac{qc_{(n,d)}}{c_{(n,d)}},~~~\text{efficiency}_{(n,d)} = \frac{qt_{(n,d)}}{t_{(n,d)}},\eqno{(8)}$$
where $qc$ and $qt$ stand for the gate cost and execution time of QISKit's algorithm, and $c$ and $t$ indicate those of our method.
Fig.7 shows that in all cases we executed, our algorithm can use fewer quantum gates to adjust and optimize the original circuits in less time.
In the worst case (more qubits and more circuit depth), we can use 6\% less gates and the efficiency is about 5 times;
in optimal case (more qubits and less circuit depth), we can use 63\% less gates and the efficiency is about 20 times.\\
\indent Obviously, the results are consistent with the theory:
when the number of qubits is large and the circuit depth is small, since we recursively calculate 4 layers of the solution space tree, the choice is more reliable and the performance is better;
when the number of qubits is small, the layout tends to be fully connected and our scheme does not have advantages;
and when the circuit depth is large, we will be easily trapped into the local optimum and the performance of our scheme is worse than that of the small depth.\\
\noindent\textbf{Performance in different physical layouts}\\
\indent For the four layouts we have chosen, there are also significant differences in costs of quantum gates and execution time.
In order to deal with different scales of circuits in a fair manner and avoid the statistical result being dominated by large-scale circuits, we no longer directly sum up the gate costs in different cases (as used in Table 1).
Specifically, the statistical method is as follows:
$$
\text{cost}_{l,c}=  \frac{1}{2240}[\sum_{i=1}^{2240}(\frac{c_{i}}{o_i})],~~~
\text{efficiency}_{l} = \frac{1}{2240}[\sum_{i=1}^{2240}(\frac{qt_{i}}{ot_i})].\eqno{(9)}
$$
where $l \in \{Linear,Circle,Center,Neighbour\}$, $c\in\{oc,qc\}$, $o_i$, $qc_i$ and $oc_i$ stand for the gate cost of the $i$th original circuit, the $i$th circuit adjusted by QISKit and our scheme respectively, $qt_i$ and $ot_i$ stand for the time required to compile the $i$th circuit by QISKit and our scheme respectively.\\
\indent Fig.8 (a) shows that for the \textit{central} layout, our scheme requires $1.80$ times the gate consumption of the original circuit, and the optimizing method of QISKit requires $3.68$ times; for the \textit{linear} layout, the gate cost of our scheme is $2.28$ times as many as the original cost and the cost of QISKit is about $2.86$ times; as for the \textit{circle} and \textit{neighbour} layouts, our scheme need to use $1.77$ times and $1.60$ times the gate cost respectively, while QISKit's method need $2.05$ times and $2.01$ times. And Fig.8 (b) illustrates that for the \textit{linear}, \textit{circle} and \textit{neighbour} layouts, our scheme is about 4 times faster than QISKit; as for the \textit{central} layout, the efficiency of our schemes is about 17.3 times as fast as QISKit's method.
\begin{figure}[!htbp]
    \begin{minipage}[t]{0.45\linewidth}
     \centering
 	  \includegraphics[width=\textwidth]{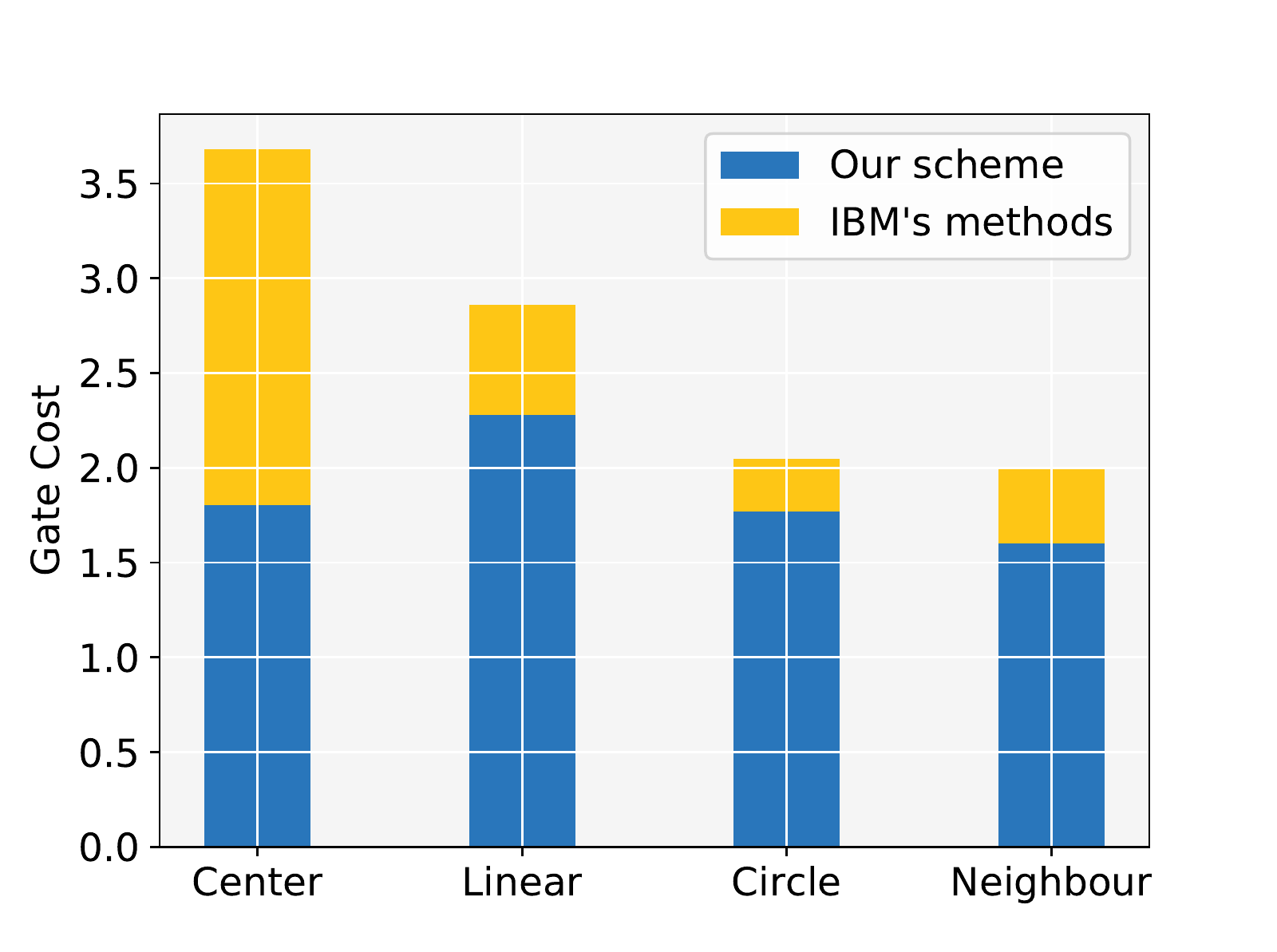}
 	  \caption*{(a) Costs of four layouts}
    \end{minipage}\hfill
 	\begin{minipage}[t]{0.45\linewidth}
     \centering
 	  \includegraphics[width=\textwidth]{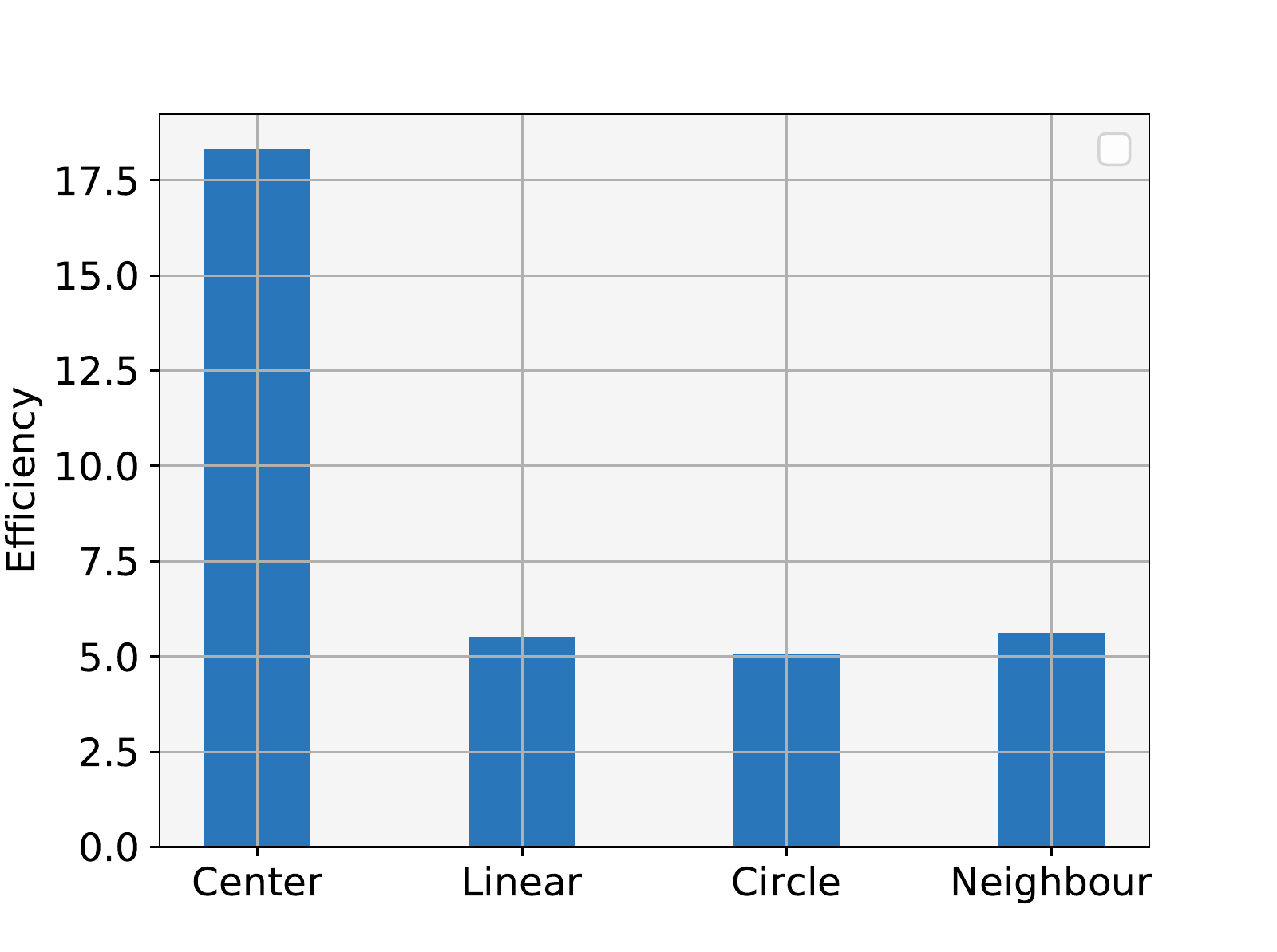}
 	  \caption*{(b) Efficiency of four layouts}
    \end{minipage}
    \caption{Experimental Results}
\end{figure}


\section{Conclusions and Future Research}
Considering the cost of physical implement, layouts of most existing quantum chips are not fully connected, which sets additional barriers for implementing quantum algorithms and programming quantum programs.
Therefore, a better approach is to automate the task of adjusting and optimizing quantum programs according to any given layout by the compiler of quantum computer.
We propose a general optimizing scheme to accomplish the task by adding additional logic gates, exchanging qubits in the quantum register and merging single-qubit gates.
Compared with QISKit's optimizing method, the quantum gates consumed by our scheme is $74.7\%$ and the execution time is only $12.9\%$ overall. For circuits with more qubits and less circuit depth, this advantage is more obvious.
In addition, several common connected graphs (linear, central, neighboring and circular) are compared as well. In these four cases, our scheme has advantages. Especially for the central layout, we can use only $49\%$ gates and $5.8\%$ execution time of QISKit's optimizing algorithm to adjust and optimize the original quantum circuits.\\
\textbf{Future Research}\\
\indent In our scheme, we often use the idea of greedy algorithm to make a choice when the circuit depth of the quantum program is deep. But the experimental results in section 4 show that we made wrong choices in some cases, and got trapped in the local optimal solution. If we can find more equitable selection criteria or even calculate the global optimal solution, we will further reduce the consumption of additional logic gates. \\
\indent In addition, a high precision floating-point calculation is needed in the combination of single-qubit logic gates, which takes up about 70\% of the total compile time. Whether we can find more efficient merging methods is a problem worth of consideration. In order to further evaluate different physical layouts, we also plan to discuss with the R\&D teams of actual quantum chips to combine the actual overhead needed to design different layouts and the expense of the software level.




\bibliographystyle{unsrt}
\bibliography{refer}

\begin{thebibliography}{10}

\bibitem{simon1997power}
Daniel~R Simon.
\newblock On the power of quantum computation.
\newblock {\em SIAM journal on computing}, 26(5):1474--1483, 1997.

\bibitem{shor1999polynomial}
Peter~W Shor.
\newblock Polynomial-time algorithms for prime factorization and discrete
  logarithms on a quantum computer.
\newblock {\em SIAM review}, 41(2):303--332, 1999.

\bibitem{grover1996fast}
Lov~K Grover.
\newblock A fast quantum mechanical algorithm for database search.
\newblock In {\em Proceedings of the twenty-eighth annual ACM symposium on
  Theory of computing}, pages 212--219. ACM, 1996.

\bibitem{cheung2007translation}
Donny Cheung, Dmitri Maslov, and Simone Severini.
\newblock Translation techniques between quantum circuit architectures.

\bibitem{linke2017experimental}
Norbert~M Linke, Dmitri Maslov, Martin Roetteler, Shantanu Debnath, Caroline
  Figgatt, Kevin~A Landsman, Kenneth Wright, and Christopher Monroe.
\newblock Experimental comparison of two quantum computing architectures.
\newblock {\em Proceedings of the National Academy of Sciences}, page
  201618020, 2017.

\bibitem{ibmbi}
The backend information of ibm quantum cloud.
\newblock \url{https://github.com/QISKit/qiskit-backend-information/}.

\bibitem{zhong2016emulating}
YP~Zhong, D~Xu, P~Wang, C~Song, QJ~Guo, WX~Liu, K~Xu, BX~Xia, C-Y Lu, Siyuan
  Han, et~al.
\newblock Emulating anyonic fractional statistical behavior in a
  superconducting quantum circuit.
\newblock {\em Physical review letters}, 117(11):110501, 2016.

\bibitem{alibaba}
The url of alibaba's quantum cloud platform.
\newblock \url{http://quantumcomputer.ac.cn/index.html}.

\bibitem{xin2017nmrcloudq}
Tao Xin, Shilin Huang, Sirui Lu, Keren Li, Zhihuang Luo, Zhangqi Yin, Jun Li,
  Dawei Lu, Guilu Long, and Bei Zeng.
\newblock Nmrcloudq: a quantum cloud experience on a nuclear magnetic resonance
  quantum computer.
\newblock {\em Science Bulletin}, 2017.

\bibitem{zeh1970interpretation}
H~Dieter Zeh.
\newblock On the interpretation of measurement in quantum theory.
\newblock {\em Foundations of Physics}, 1(1):69--76, 1970.

\bibitem{divincenzo2000physical}
David~P DiVincenzo et~al.
\newblock The physical implementation of quantum computation.
\newblock {\em arXiv preprint quant-ph/0002077}, 2000.

\bibitem{yao1993quantum}
A~Chi-Chih Yao.
\newblock Quantum circuit complexity.
\newblock In {\em Foundations of Computer Science, 1993. Proceedings., 34th
  Annual Symposium on}, pages 352--361. IEEE, 1993.

\bibitem{qiskit}
Qiskit python api.
\newblock \url{https://qiskit.org/}.

\bibitem{qiskitc}
Qiskit developer challenge.
\newblock \url{https://qx-awards.mybluemix.net/}.

\bibitem{pednault2017breaking}
Edwin Pednault, John~A Gunnels, Giacomo Nannicini, Lior Horesh, Thomas
  Magerlein, Edgar Solomonik, and Robert Wisnieff.
\newblock Breaking the 49-qubit barrier in the simulation of quantum circuits.
\newblock {\em arXiv preprint arXiv:1710.05867}, 2017.

\bibitem{cross2017open}
Andrew~W Cross, Lev~S Bishop, John~A Smolin, and Jay~M Gambetta.
\newblock Open quantum assembly language.
\newblock {\em arXiv preprint arXiv:1707.03429}, 2017.

\bibitem{svore2006layered}
Krysta~M Svore, Alfred~V Aho, Andrew~W Cross, Isaac Chuang, and Igor~L Markov.
\newblock A layered software architecture for quantum computing design tools.
\newblock {\em Computer}, 39(1):74--83, 2006.

\bibitem{Barenco1995Elementary}
A~Barenco, C.~H. Bennett, R~Cleve, D.~P. Divincenzo, N~Margolus, P~Shor,
  T~Sleator, J.~A. Smolin, and H~Weinfurter.
\newblock Elementary gates for quantum computation.
\newblock {\em Physical Review A}, 52(5):3457, 1995.

\bibitem{nielsen2002quantum}
Michael~A Nielsen and Isaac Chuang.
\newblock Quantum computation and quantum information, 2002.

\bibitem{u3u3}
The code of merging two single gates.
\newblock
  \url{https://github.com/QISKit/qiskit-sdk-py/blob/master/qiskit/mapper/_mapping.py}.

\bibitem{iachello2006lie}
Francesco Iachello.
\newblock {\em Lie algebras and applications}, volume 708.
\newblock Springer, 2006.

\bibitem{vatan2004optimal}
Farrokh Vatan and Colin Williams.
\newblock Optimal quantum circuits for general two-qubit gates.
\newblock {\em Physical Review A}, 69(3):032315, 2004.

\end{thebibliography}
\end{document}